\documentclass[aps,showpacs,superscriptaddress,floatfix]{revtex4}

\usepackage{graphicx}
\usepackage{amsmath}
\usepackage{amssymb}

\begin{document}

\title{Photostatistics Reconstruction via Loop Detector Signatures}

\author{J. G. Webb}
\affiliation{Centre for Quantum Computer Technology, School of Information
Technology and Electrical Engineering, University College, The
University of New South Wales, Canberra, ACT, 2600}
\email{james.webb@adfa.edu.au}

\author{E. H. Huntington}
\affiliation{Centre for Quantum Computer Technology, School of Information
Technology and Electrical Engineering, University College, The
University of New South Wales, Canberra, ACT, 2600}

\begin{abstract}
Photon-number resolving detectors are a fundamental building-block of optical quantum information processing protocols.  A loop detector, combined with appropriate statistical processing, can be used to convert a binary on/off photon counter into a photon-number-resolving detector.  Here we describe the idea of a signature of photon-counts, which may be used to more robustly reconstruct the photon number distribution of a quantum state.  The methodology is applied experimentally in a 9-port loop detector operating at a telecommunications wavelength and compared directly to the approach whereby only the number of photon-counts is used to reconstruct the input distribution.  The signature approach is shown to be more robust against calibration errors, exhibit reduced statistical uncertainty, and reduced reliance on {\em a-priori} assumptions about the input state.
\end{abstract}


\maketitle
\vspace{10 mm}

\section{Introduction}

One of the most promising applications of quantum science lies in the field of quantum information processing \cite{nielsen}.  Quantum optics provides an extremely convenient
 test bed for quantum information ideas because quantum effects dominate even
 at room temperatures and non-classical states are
 relatively easy to achieve in the  laboratory.  This has been most impressively illustrated with recent small-scale experimental demonstrations of quantum algorithms \cite{OPWRB03,LWLBJGW}.

Photon-number resolving detectors, as well as the ability to characterise quantum optical states using such detectors, are a key element in the further development of quantum optical information processing systems \cite{LOQCreview,QOqip}.    Due to the low energies carried by photons, contemporary detectors are incapable of resolving photon number with high fidelity.  Typically linear PIN photodiodes used as direct detectors of the optical field are limited to mesoscopic photon-number regimes \cite{BondaniArXiv}.  Recent results show promise at telecommunications wavelengths using liquid-He cooled diodes \cite{Fujiwara}, bolometric devices \cite{supercond2} and super-conducting nano-structures \cite{supercond1} although such solutions are not yet widespread due to their inherent cost and/or cryogenic cooling complexities.

Consequently, use is commonly made of single photon detector modules (SPDMs), based on avalanche photo-diodes (APDs).  A number of techniques have been developed for approximating number-resolving detection using these inherently non-number-resolving detectors \cite{Kok01,PhotonChopPRL,Bartlett02,Rohde05,AchillesJMO,BanaszekOptLett03,FitchPRA,Pereira07,Kim99,Takeuchi99}, all of which are variations on the idea that the optical field is distributed across multiple modes which are measured independently. The experiment is arranged so that the probability of any given mode being populated by more than one photon approaches zero.  Thus in the absence of dark counts, afterpulsing or imperfect quantum efficiency, the sum of the detection events across the modes closely approximates the number of photons in the incident state.

In the case of the ideal $N$-port multiplexer with ideal binary photon detectors, the maximum number of photons able to be resolved is equal to the number of output modes.  This limit arises due to excess input photons resulting in $>1$ photons in the output modes, which are not distinguishable from the single photon case.  In a realistic detector, however, the presence of loss means that $n$ input photons may result in $m$ clicks where $n \ge m$.  If $m\le N$ and $n > N$ then it can be seen that the loss gives rise to a useable increase in the detector's dynamic range.  In general though, the accuracy of the reconstruction degrades and the statistical errors increase with increasing amounts of loss.  Similarly, fragile non-classical effects are washed out by increased losses, as quantified in \cite{ZambraPRA}.  This situation is further complicated by the presence of dark counts and afterpulsing.

There have been many demonstrated solutions to the reconstruction of the photostatistics (photon number distribution) of an input field using binary (on/off) detectors, but they may each be loosely grouped into one of three categories.  All realistic photodetectors exhibit loss, which is modelled by preceding an otherwise ideal detector with a beamsplitter of transmission ${\eta}$. The photostatistics of the input state are thus convolved with the Bernoulli (binomial) distribution arising from the stochastic photon selection process performed at the beamsplitter \cite{WallsMilburn}.

This relationship forms the basis of the first category of reconstruction techniques - those involving inversion of the Bernoulli distribution. The analytic inversion is cumbersome and requires $\eta \ge 0.5$ \cite{ArianoPRA}, an impossible constraint on N-ports with $N>2$ which is further exaggerated at telecommunications wavelengths due to the very small quantum efficiencies.  Numerical approaches such as maximum likelihood techniques permit more realistic values of $\eta$ and have been successfully applied in \cite{MogilevtsevOptComm98,ZambraIJQI}.

The second category involves reconstruction of the input density matrix, $\rho$, via conditional probabilities and appears the most common contemporary approach.  An N-port detector may be completely described by the corresponding conditional probability matrix $P(m|n)$ which quantifies the probability of observing $m$ clicks, given $n$ incident photons.  The experiments of \cite{AchillesJMO,FitchPRA,LeeJModOpt04} perform a large number of measurements of an unknown input state, and record the number of clicks obtained from each measurement.  By applying either the inverted matrix of $P(m|n)$ values to the observed ``number of click'' statistics or utilising maximum likelihood techniques, the input photon number distribution may be obtained.

Finally, as observed by \cite{RehacekPRA03}, an improvement in the accuracy of the reconstruction is possible by noting the coincidences of the $m$ detection events of the $N$ detectors.  This is equivalent to recording detection signatures for each measurement and forms the basis of our reconstruction implementation.  To date, experimental work in this third category has considered only $N=2$ distinct output modes \cite{BridaOptSpect} for a number of detection efficiencies $\eta$.  The purpose of this paper is to present experimental results for $N=9$ time-domain modes with a fixed (small) value of $\eta$ for each mode.

\section{Numerical methods}
\label{sec:numMethods}

Irrespective of the reconstruction category, the three approaches share the same goal common to all tomography problems - the calculation of the input state from a large data set of marginal distributions.  Although the analytic relationship between the measured marginal probability distributions is usually a known function of the input state, this is not the case for the inverse in general.  Even if an inverse analytic solution exists it is generally of no use as experimental and measurement uncertainties create an \emph{overcomplete} data set for which there is no unique solution for $\rho$.  The problem may thus be restated as the determination of the best \emph{estimate} $\hat{\rho}$ for $\rho$ given the measured data set.

The estimation of $\hat{\rho}$ is significantly simplified by \emph{a priori} assumptions of aspects of the statistics of the input state.  If it may be assumed that the input state consists of negligible contributions from higher order Fock states, i.e. $\langle k |\rho|k \rangle \approx 0\ \forall\ k \ge K$, then the reconstructed Hilbert space may be restricted to dimension $(K+1) \times (K+1)$.  Additional constraints such as the expected distribution (e.g. Poissonian) and physicality via $\mathrm{Tr}[\hat{\rho}] = 1$ and $\hat{\rho}_{kk} \ge 0\ \forall\ k < K$ further aid the reconstruction process.

Detection with on/off photodetectors will result in a particular pattern of detector clicks, which we call a \emph{signature}. The probability, $P_\mathrm{sig}(\vec{d})$, of the detection signature occurring whereby each of the detectors in vector $\vec{d}$ trigger, and all others do not is given by \cite{rohdeWebb}
\begin{eqnarray}
\label{eq:signatures}
P_\mathrm{sig}(\vec{d}) &=& \sum_{n_1+\dots+n_N=n} \frac{n!}{n_1! \dots n_N!} \, \prod_{i=1}^N p_c(i)^{n_i} \nonumber\\
&\times& \prod_{j\in \vec{d}} \left[p_\mathrm{dc} + (1 - p_\mathrm{dc}) [1 - p_\mathrm{loss}(j)^{n_j}] \right] \nonumber\\
&\times& \prod_{l\notin \vec{d}} \left[(1 - p_\mathrm{dc}) p_\mathrm{loss}(l)^{n_l} \right].
\end{eqnarray}

\noindent where $p_c(i)$ represents the probability of an input photon being coupled out of the $\mathrm{i}^\mathrm{th}$ port.  Similarly, $p_\mathrm{loss}(j)$ indicates the probability of photon loss associated with port $j$.  Both of these parameters are a function of the detector architecture and are detailed in \cite{rohdeWebb}.  The dark count probability $p_\mathrm{dc}$ depends upon the particular on/off detector employed.

The net probability of detecting $m$ photons, given by summing over all combinations of detection signatures where $|\vec{d}|=m$, for $|\vec{d}|$ is total number of clicks, is thus
\begin{equation}
\label{eq:pmn}
P(m|n) = \sum_{|\vec{d}|=m}P_\mathrm{sig}(\vec{d}).
\end{equation}

It is worth noting that $P_\mathrm{sig}(\vec{d})$ is not necessarily the same for all $\vec{d}$ where $|\vec{d}|=m$. That is, different signatures corresponding to the same measured number of photons needn't have equal probabilities of occurring.

For the three scenarios above, the measured data set is reduced to a vector of observed probabilities for a detection in the relevant basis.  In the binomial reconstruction approach, this corresponds to the probability of a click $\hat{p}_\mathrm{click}$ for each of the experimental values of detection efficiency $\eta_x$.  The probability $p_\mathrm{out}(m)$ of obtaining $m$ output photons depends on the $m^\mathrm{th}$ \emph{and subsequent} terms of the input state density matrix $\rho$. i.e.

\begin{equation}
\label{eq:pPhoton}
p_\mathrm{out}^{(m)} = \sum_{i = m}^{\infty}\binom{i}{m}\eta^m(1-\eta)^{i-m}\rho_{ii}.
\end{equation}

Incorporating the independent dark count statistics and assuming a binary on/off detector (typically an SPDM) with a linear response to photon number, Eq.\ \ref{eq:pPhoton} may be integrated over the truncated Hilbert space to give

\begin{equation}
\label{eq:pClickBinom}
p_\mathrm{click}(\eta, \rho) = p_\mathrm{dc} + (1-p_\mathrm{dc})\sum_{n=1}^k\sum_{j\ge n}^k\binom{j}{n}\eta^n(1-\eta)^{j-n}\rho_{nn}
\end{equation}

Corresponding expressions can be obtained for the photon counting and signature approaches from Eqs.\ \ref{eq:signatures} and  \ref{eq:pmn} respectively, i.e.

\begin{equation}
\label{eq:pmnTomog}
p(m, \rho) = \sum_{|\vec{d}|=m}p_\mathrm{sig}(\vec{d}, \rho)
\end{equation}

where

\begin{eqnarray}
\label{eq:signaturesTomog}
p_\mathrm{sig}(\vec{d}, \rho) &=& \sum_{n=0}^k \rho_{nn}\Bigg[ \sum_{n_1+\dots+n_N=n} \frac{n!}{n_1! \dots n_N!} \, \prod_{i=1}^N p_c(i)^{n_i} \nonumber\\
&\times& \prod_{j\in \vec{d}} \left[p_\mathrm{dc} + (1 - p_\mathrm{dc}) [1 - p_\mathrm{loss}(j)^{n_j}] \right] \nonumber\\
&\times& \prod_{l\notin \vec{d}} \left[(1 - p_\mathrm{dc}) p_\mathrm{loss}(l)^{n_l} \right] \Bigg]
\end{eqnarray}

As stated earlier, these expressions allow the calculation of the expected statistics for a known density matrix.  To reconstruct the unknown state, the statistical distance between the predicted and measured data sets is quantified.  A correction is thus determined and applied to the current state prediction to create a new estimate.  This process is repeated until either the desired degree of convergence between the datasets is achieved or the incremental changes to the estimated state fall below an arbitrary threshold.

The determination of the optimal means of quantifying the statistical distance between the data sets is an area of ongoing research.  Similarly, the calculation of the corresponding amendment to the estimated state is also a non-trivial exercise.  A comparison of the popular contemporary methods of maximum likelihood and maximum entropy approaches are discussed in the context of the inversion of photostatistics data in \cite{HradilRehacek}.

For our purposes, we quantify the distance between the theoretical and measured data sets via the mean square error between the distributions used as a measure of convergence by \cite{ZambraPRL}.  If $p$ and $\hat{p}$ are the calculated and observed marginal probability distributions in the basis applicable to the reconstruction category, and $S$ is the total data set size,

\begin{equation}
\label{eq:mseTomog}
\epsilon = \sum_i^S (p_i - \hat{p}_i)^2
\end{equation}

The problem may hence be expressed as a multidimensional numerical minimisation problem, where the goal is the minimisation of $\epsilon$ by appropriate selection of the $(K+1)$ diagonal coefficients of $\rho$.  For this task, the simplex method of Nelder and Mead \cite{NelderMead} was chosen for ease of implementation \cite{NumericalRecipes} and the requirement for evaluation of $\epsilon$ only, and not its derivatives.  Although convergence of the algorithm to a global minimum is not guaranteed \cite{SimplexConverge}, in practice its convergence was found to be fast and entirely adequate for a wide range of expressions for $\epsilon$.

\section{1550 nm loop detector}

\begin{figure}[!htb]
\begin{center}
    \includegraphics[width=0.8\textwidth]{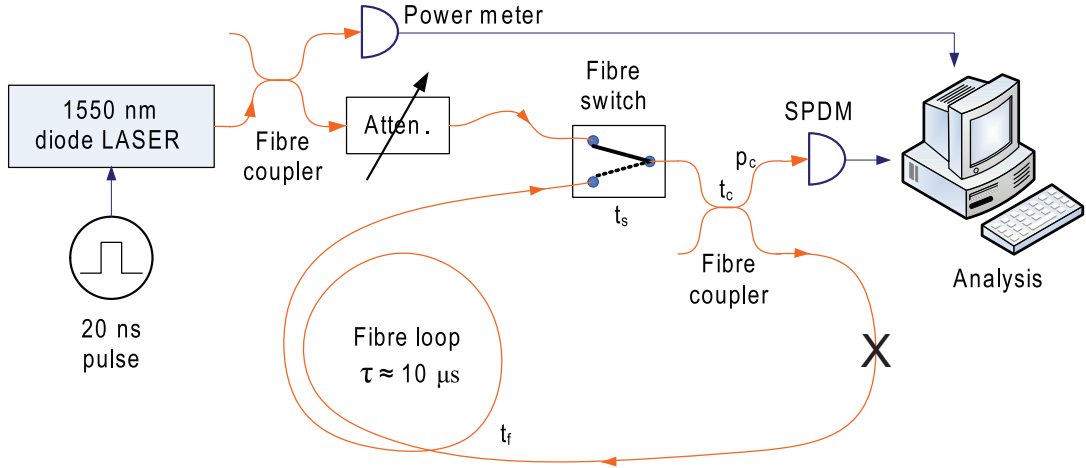}
    \caption{Experimental configuration of the loop detector and diode laser source of pulsed coherent states. The fibre loop may be broken (disconnected) at point 'X' for characterisation purposes as described in the text.}
    \label{fig:loopExpt}
\end{center}
\end{figure}

The TDM loop detector architecture of \cite{BanaszekOptLett03} was selected because of its inherent flexibility in permitting a variable number $N$ of output modes.  The architecture is shown in Fig.\ \ref{fig:loopExpt}.

The experiment allows the various reconstruction methods to be assessed against the Poissonian photostatistics of the coherent output of the diode laser.  The laser is biased below threshold, and pulsed briefly into operation by the application of a variable amplitude 20 ns pulse.  95\% of the laser output is directed to a NIST-traceable Newport 3227 power meter for calibration/renormalisation purposes.  The remaining 5\% is then transferred via a series of calibrated fibre attenuators providing $\sim70$ dB of attenuation into the loop detector.

The loop detector comprises the input switch, output coupler and fibre delay loop.  The switch was provided by Civcom Inc. and provides a remarkably low insertion loss (1.2 dB) combined with fast (270 ns) switching speed.  As the fibre provides $10.044\ \mu$s of delay, the switch thus has adequate time to settle in the alternative state (shown by the dotted line of Fig.\ \ref{fig:loopExpt}) after admitting the pulse to the detector.  The loop consists of 2 km of Corning SMF-28 fibre, exhibiting only 0.8 dB loss at 1550 nm, with each coupler exhibiting 0.5dB of loss.  The long delay allows for the recovery of the APD used in the SPDM following a gate pulse, in order to minimise the detrimental effects of afterpulsing. A portion of the beam is tapped off from the circulating field via the fibre coupler at each pass and directed onto an id-Quantique id-200 SPDM operating with a 20 ns gate window and a 100 kHz clock rate.  This detector had a quantum efficiency of 10 \% and a dark count probability per gate $p_\mathrm{dc}=9.6 \times 10^{-4}$.

$N=14$ bins was selected with a coupling coefficient $p_c = 0.1$, yielding a possible truncation error of $23$\% \cite{rohdeWebb}.  In practice, this error is not observed due to the excess loop losses preventing photons making it into the last few bins for the range of input intensities employed.  Combined with the $\eta_d = 10$\% quantum efficiency of the id-Quantique id-200 SPDM used, the equivalent detection efficiencies $\eta_i$ for each time bin are shown in Table\ \ref{table:etaVals}.

\begin{table}[!htb]
\begin{center}
\caption{Detection efficiencies corresponding to loop TDM bins.}
\begin{tabular}{|c|c|c|c|c|c|c|c|}
\hline
Bin   & 1 & 2 & 3 & 4 & 5 & 6 & 7\\
$\eta$& 6.83e-3 & 3.58e-3 & 1.88e-3 & 9.91e-4 & 5.22e-4 & 2.75e-4 & 1.45e-4\\
\hline
Bin & 8 & 9 & 10 & 11 & 12 & 13 & 14 \\
$\eta$& 7.61e-5 & 4.01e-5 & 2.11e-5 & 1.11e-5 & 5.84e-6 & 3.08e-6 & 1.62e-6 \\
\hline
\end{tabular}
\label{table:etaVals}
\end{center}
\end{table}

Due to the long delay employed, the loop path length is much greater than the coherence length of the laser.  Consequently, the tapped field is uncorrelated with the circulating component within the loop, and any quantum correlations between successive time bins are destroyed.  This allows the loop data be considered mathematically classical and photons to be treated as discrete particles.

Accurate timing of the experiment is critical to ensure the APD gate window that defines each time bin is correctly aligned with the pulse circulating within the loop.  Control of the timing is achieved via a synchronous finite state machine (FSM) constructed within a field programmable gate array (FPGA).  A measurement \emph{cycle} consisting of the generation of an optical pulse, recording 14 passes of the loop and transferring the data to file for post processing requires $376 \mu$s, allowing 2660 distinct measurements per second to be performed.

\begin{figure}[!htb]
\begin{minipage}{0.45\linewidth}
\raisebox{-1ex}{\makebox[\textwidth][l]{\textbf{a)}}}
\centering
\includegraphics[width=\textwidth]{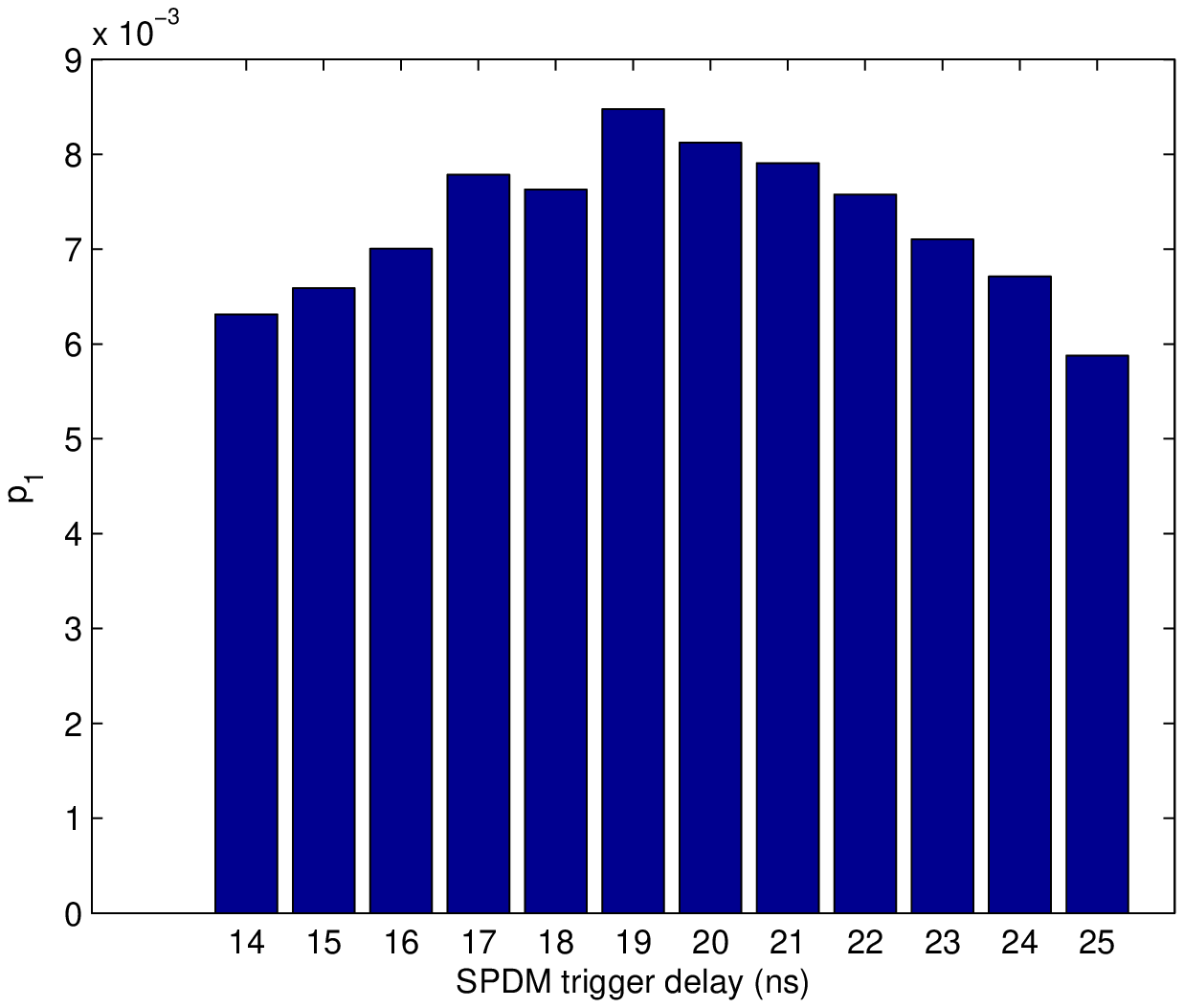}
\end{minipage}
\begin{minipage}{0.45\linewidth}
\raisebox{-1ex}{\makebox[\textwidth][l]{\textbf{b)}}}
\centering
\includegraphics[width=\textwidth]{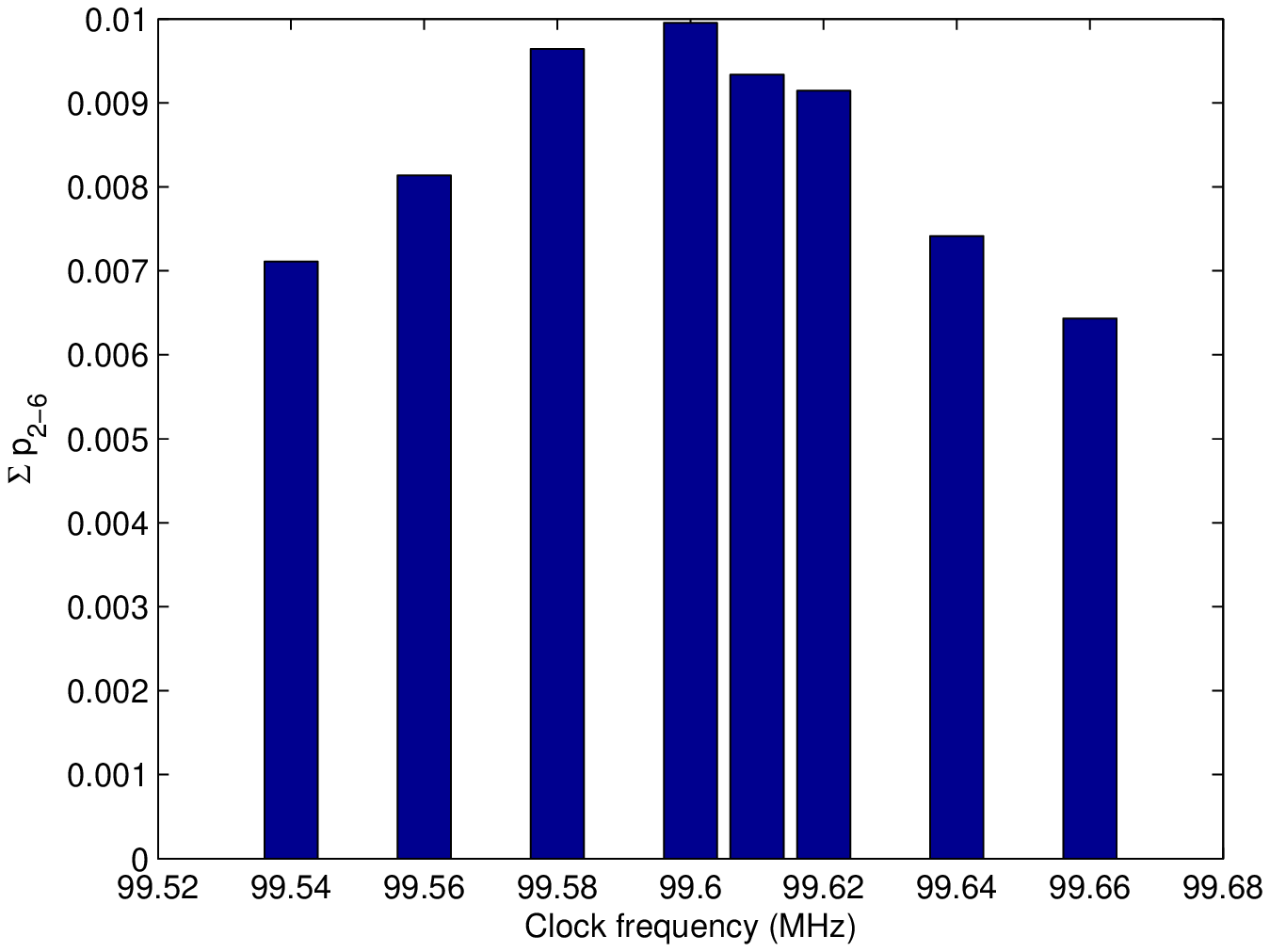}
\end{minipage}
\caption{Determination of a) optimum SPDM trigger delay and b) clock frequency for the loop detector.  The SPDM trigger delay is chosen to maximise the detection probability of the first time-bin.  The clock frequency is chosen to maximise $\sum_{i=2}^5 p_i$.}
\label{fig:trigDelay}
\end{figure}

\begin{figure}[!htb]
\begin{center}
    \includegraphics[width=0.5\textwidth]{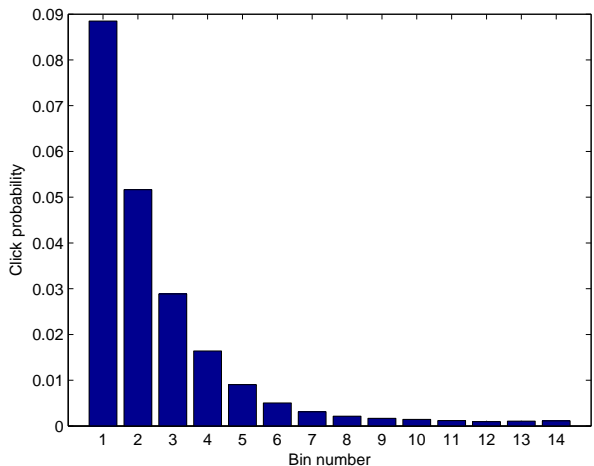}
    \caption{Raw time bin detection (click) probabilities for $\eta_d = 10$\%, $\bar{n}=8.1$ photons/pulse.}
    \label{fig:detProbs}
\end{center}
\end{figure}

Compensation for the multitude of electro-optic delays within the experiment is achieved by adjusting the triggering time of the SPDM and the clock frequency of the FSM.  Fig.\ \ref{fig:trigDelay}  indicates the frequency of detection events in the first (and subsequent) time bins as a function of time delay and clock frequency respectively.  The final timing parameters chosen thus allow the detection efficiencies to approach those listed in Table\ \ref{table:etaVals}.  Fig.\ \ref{fig:detProbs} illustrates the resulting exponentially decaying click probabilities obtained for each bin.

In practice, it was noted that very few valid detection events were observed in the later time bins.  This is as expected due to the vanishingly small effective detection efficiencies.  Consequently, the analysis was terminated at $N=9$ time bins.  To compensate for the induced truncation error, a non-experimentally observable $10^\mathrm{th}$ bin was re-introduced to the expression of Eq.\ \ref{eq:signaturesTomog} to serve as a ``catch all'' for photons normally lost to the analysis.

To accurately model the 9 time-bin experimental data, Eq.\ \ref{eq:signaturesTomog} is thus rewritten

\begin{equation}
\label{eq:psigTrunc}
p_\mathrm{sig}(\vec{d}, \rho)_{N=9} \approx p_\mathrm{sig}(\vec{[d\ 0]}, \rho)_{N=10} + p_\mathrm{sig}(\vec{[d\ 1]}, \rho)_{N=10}
\end{equation}

\noindent and the $10^\mathrm{th}$ element of the coupling and loss vectors are such that $p_c(10) = 1 - \sum_{i=1}^9 p_c(i)$ and $p_\mathrm{loss}(10) = 0$.

As a test of validity, this ensures
\begin{equation}
\label{eq:psigTruncOK}
\sum_{\vec{d}} p_\mathrm{sig}(\vec{d}, \rho)_{N=9} = 1\ \forall\ n,\rho
\end{equation}

\section{Characterisation}
\label{sec:loopChar}

\begin{figure}[!htb]
\begin{center}
    \includegraphics[width=0.6\textwidth]{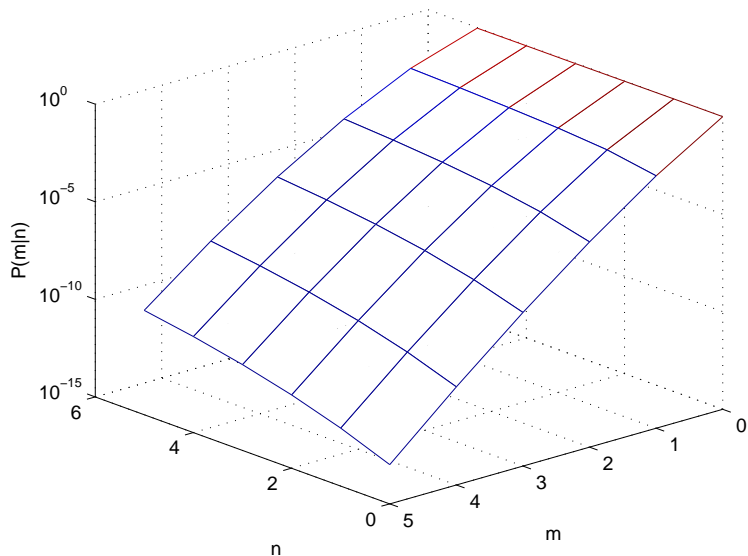}
    \caption{Characteristic conditional probabilities $P(m|n)$ for the 1550 nm loop detector.  Note the logarithmic vertical axis used to accommodate the rapidly diminishing higher order terms.}
    \label{fig:condProbs}
\end{center}
\end{figure}

In addition to the bin efficiencies listed in Table\ \ref{table:etaVals}, our detector can be completely characterised by its conditional probability matrix.  The first few terms are listed in Table\ \ref{table:pmnVals} and plotted graphically in Fig.\ \ref{fig:condProbs}.

\begin{table}[!htb]
\begin{center}
\caption{$P(m|n)$ for loop detector.}
\begin{tabular}{|c|c|c|c|c|c|c|}
\hline
$m\ \backslash\ n$ & 0 & 1 & 2 & 3 & 4 & 5 \\
\hline
0 & 0.9913 & 0.9776 & 0.9640 & 0.9506 & 0.9374 & 0.9244 \\
\hline
1 & 0.0086 & 0.0222 & 0.0356 & 0.0487 & 0.0614 & 0.0738 \\
\hline
2 & 3.29e-5 & 1.39e-4 & 3.70e-4 & 7.24e-4 & 1.19e-3 & 1.78e-3 \\
\hline
3 & 7.39e-8 & 4.29e-7 & 1.64e-6 & 4.45e-6 & 9.59e-6 & 1.77e-5 \\
\hline
4 & 1.06e-10 & 7.90e-10 & 3.95e-9 & 1.39e-8 & 3.79e-8 & 8.56e-8 \\
\hline
5 & 1.02e-13 & 9.24e-13 & 5.71e-12 & 2.50e-11 & 8.30e-11 & 2.23e-10 \\
\hline
\end{tabular}
\label{table:pmnVals}
\end{center}
\end{table}

Immediately apparent is the departure from the ideal $P(m|n)=\delta_{m,n}$, arising as a direct consequence of the loop losses and the poor InGaAs APD efficiency.  The quantum efficiency of the id-200 can be doubled at the cost of an order of magnitude increase in dark counts, i.e. at $\eta_d = 10$\%, $p_\mathrm{dc} = 9.66\times 10^{-4}$, while at $\eta_d = 25$\%, $p_\mathrm{dc} = 2.75\times 10^{-2}$.  However, dark counts adversely affect fidelity to a greater extent than reduced quantum efficiency \cite{rohdeWebb}.  Consequently, the lower of the id-200 detection efficiencies were used.

\begin{figure}[!htb]
\begin{minipage}{0.45\linewidth}
\raisebox{-1ex}{\makebox[\textwidth][l]{\textbf{a)}}}
\centering
\includegraphics[width=\textwidth]{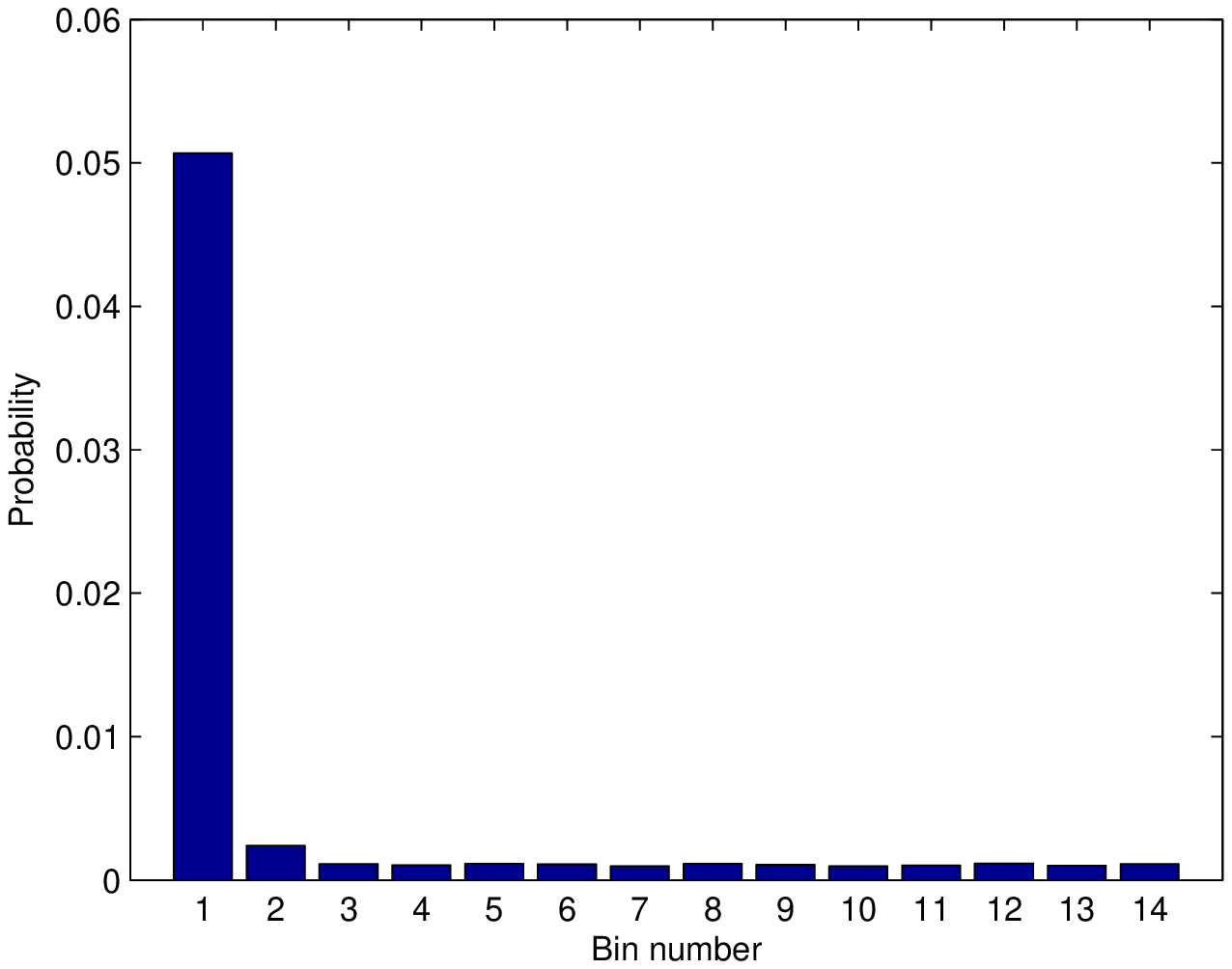}
\end{minipage}
\begin{minipage}{0.45\linewidth}
\raisebox{-1ex}{\makebox[\textwidth][l]{\textbf{b)}}}
\centering
\includegraphics[width=\textwidth]{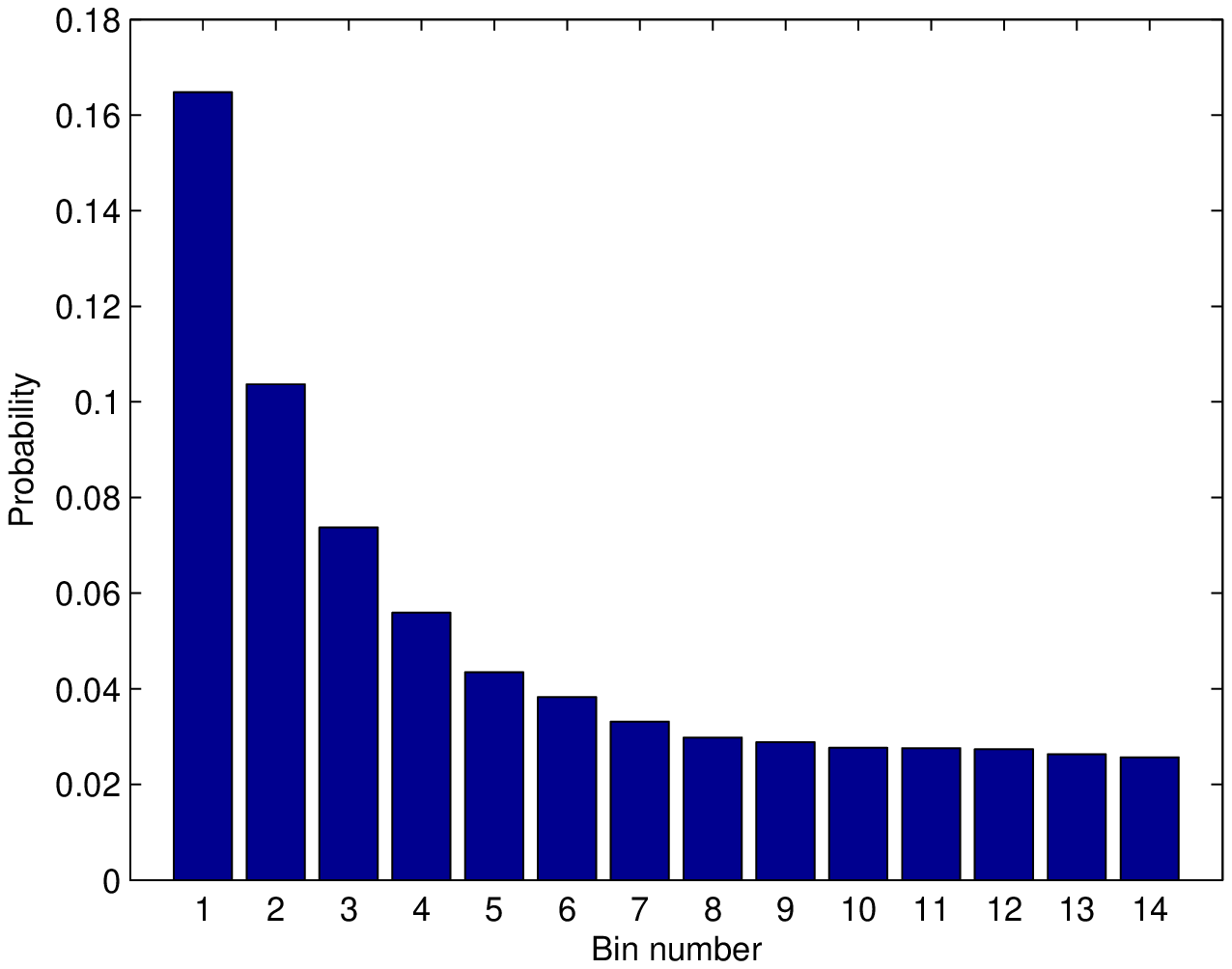}
\end{minipage}
\caption{Measured time-bin detection probabilities for a) $\eta_d = 10$\% and b) $\eta_d = 25$\%, with the loop open-circuit.}
\label{fig:loopOpen25}
\end{figure}

The remnant dark count ``noise floor'' is visible at the exponential tail of Fig.\ \ref{fig:detProbs} count probabilities.  Fig.\ \ref{fig:loopOpen25} provides further insight into the limitations of the photodetection process, illustrating the effect of disconnecting the loop at point X as shown in Fig.\ \ref{fig:loopExpt} for $\eta_d = 10$\% and $\eta_d = 25$\% respectively.  The non-zero detection probabilities in bins $\ge 2$ are the result of dark counts and \emph{afterpulsing}.

Afterpulsing describes the mechanism by which charges trapped within defects of the APD semiconductor give rise to spontaneous avalanche events during the next gate pulse, even in the absence of a detected photon.  Unfortunately, since avalanches induced by a valid photon detection event are indistinguishable from APD defect/intrinsic dark count avalanche events, afterpulsing can artificially inflate the effective dark noise floor of many subsequent detector time bins.  For fixed APD temperature and gating period, the afterpulsing probability scales non-linearly with $\eta_d$ \cite{RibordyApplOpt}.  This effect is clearly shown in Fig.\ \ref{fig:loopOpen25}(b), where photons are detected only during time bin 1 and afterpulsing is responsible for the exponential decay to the dark noise background, which is substantially greater than the $\eta_d=10$\% case of Fig.\ \ref{fig:loopOpen25}(a).

\begin{figure}[!htb]
\begin{center}
    \includegraphics[width=0.5\textwidth]{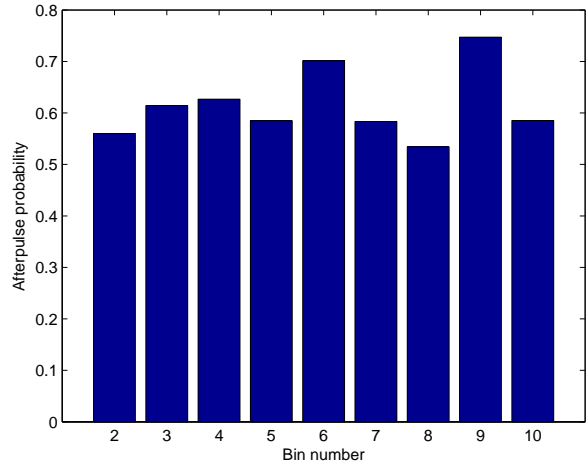}
    \caption{Measurement of afterpulsing from time series data of Fig.\ \ref{fig:loopOpen25} ($\eta_d = 25$\%).}
    \label{fig:afterpulsingProb}
\end{center}
\end{figure}

In order to quantify the effects of afterpulsing, the asymptotic background value dark noise was assumed to be independent of detection events and was subtracted away from the open loop data. Fig.\ \ref{fig:afterpulsingProb} indicates the result of comparing the probability of a click in each bin to the one before for the data of Fig.\ \ref{fig:loopOpen25}(b).  A significant 60\% of detection events in subsequent time bins are thus observed to result solely from afterpulsing effects.  The approximately constant envelope of probabilities suggests the simple ``regenerative afterpulsing'' model described in \cite{afterpulsingRevSciInst} adequately models our experiment.  Assuming $p_\mathrm{dc} \ll 1$,  the detected $p_\mathrm{click}$ data in bin $i$ may thus be corrected to first order by subtracting a portion of the counts $p_a$ detected in a prior time bin, i.e.

\begin{displaymath}
\label{eq:afterpulseCorr}
\hat{p}_\mathrm{click}(i) \approx \left\{
\begin{array}{ll}
p_\mathrm{dc} + p_\mathrm{click}(i)[1 - p_\mathrm{dc}] & \mathrm{; i = 1}\\
p_\mathrm{click}(i) - p_\mathrm{dc} - p_a p_\mathrm{click}(i-1) & \mathrm{; i \ge 2}\\
\end{array} \right.
\end{displaymath}

Note that this correction aids reconstruction via the binomial method only, as integration of the comprehensive afterpulsing model of \cite{afterpulsingRevSciInst} into the expressions of Eqs.\ \ref{eq:psigTrunc} and\ \ref{eq:pmnTomog} is non-trivial.

\begin{figure}[!htb]
\begin{minipage}{0.45\linewidth}
\raisebox{-1ex}{\makebox[\textwidth][l]{\textbf{a)}}}
\centering
\includegraphics[width=\textwidth]{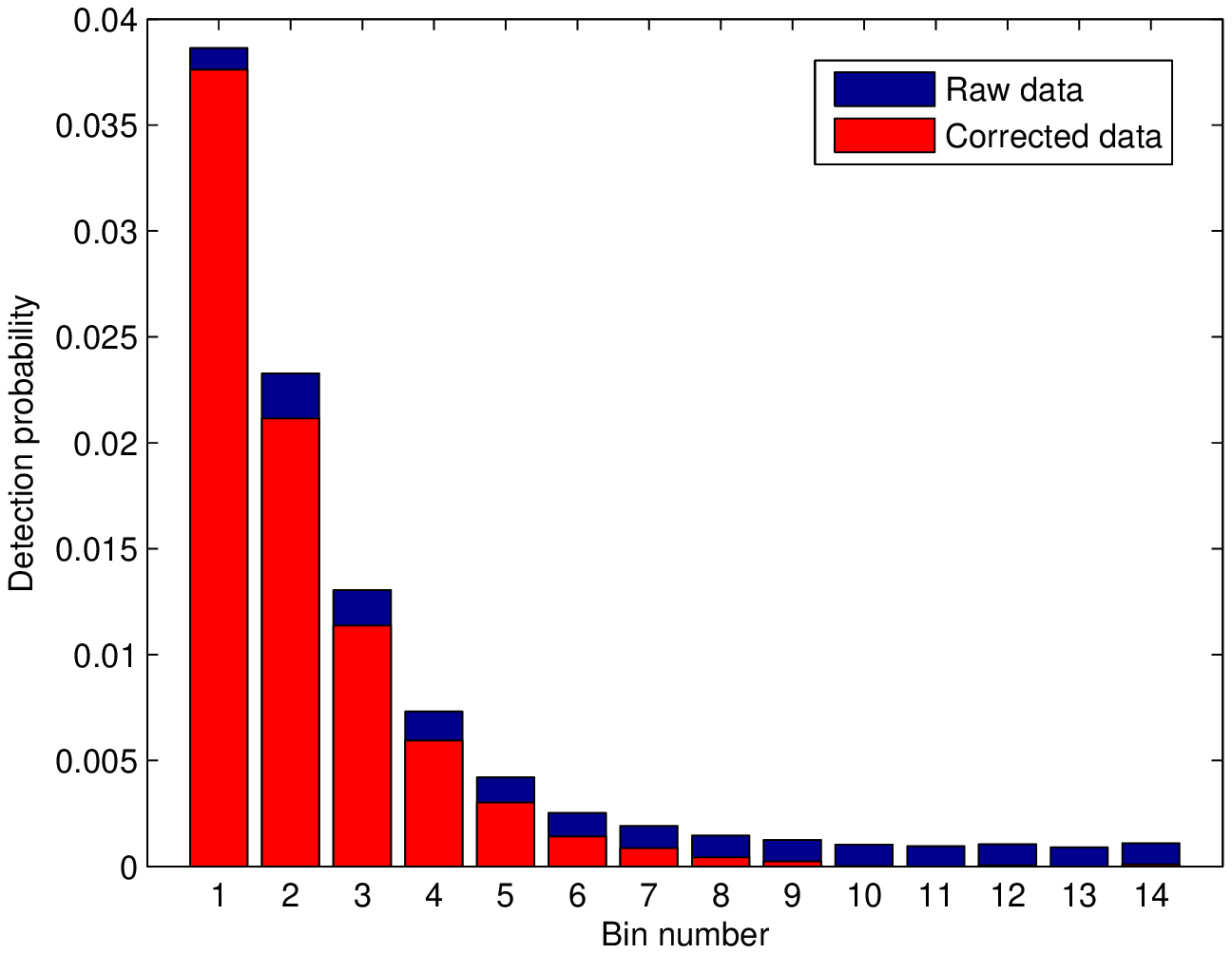}
\end{minipage}
\begin{minipage}{0.45\linewidth}
\raisebox{-1ex}{\makebox[\textwidth][l]{\textbf{b)}}}
\centering
\includegraphics[width=\textwidth]{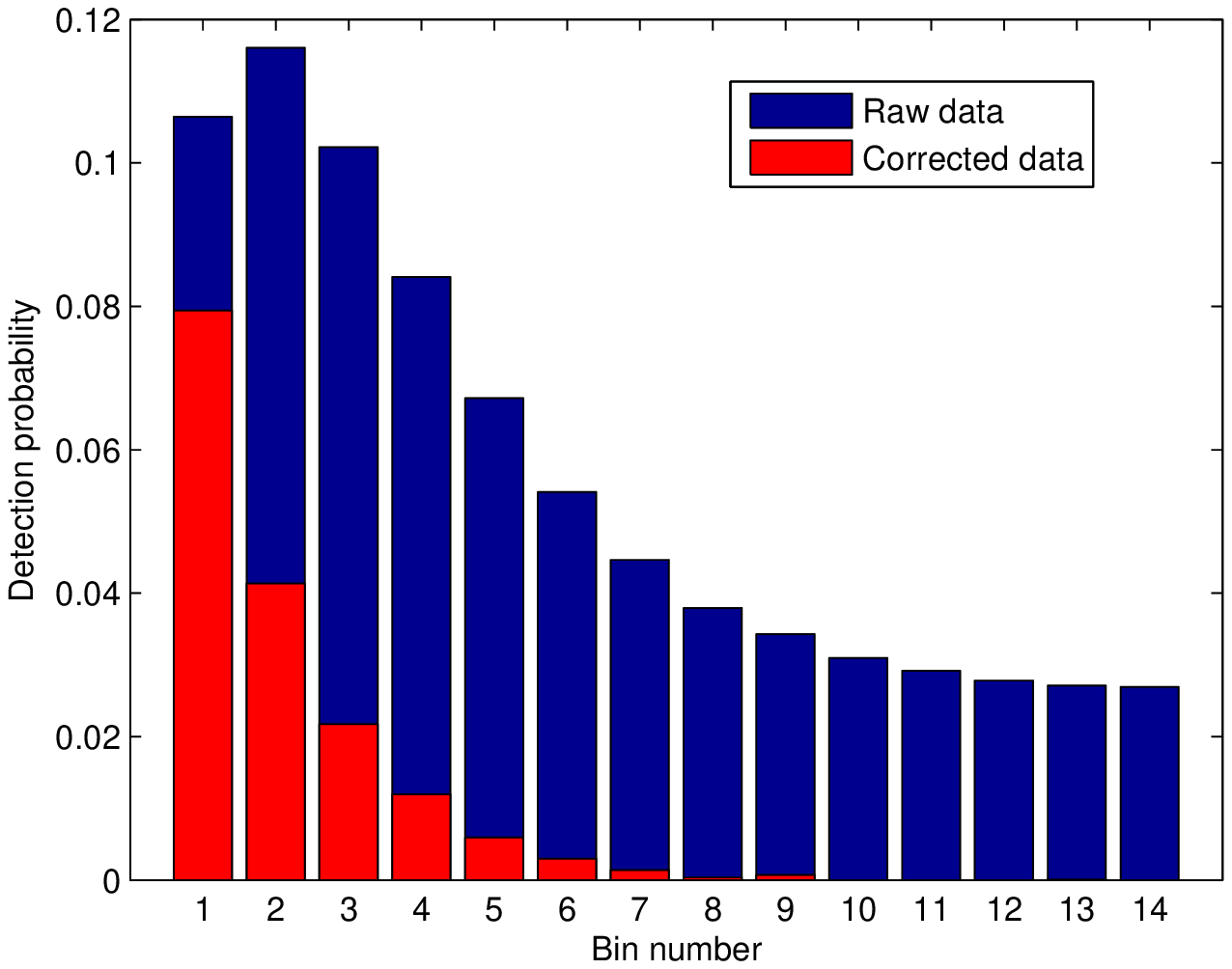}
\end{minipage}
\caption{Raw and corrected time bin detection probabilities for a) $\eta_d = 10$\% and b) $\eta_d = 25$\% with $\bar{n}=3.25$ photons/pulse.}
\label{fig:correctedDetProbs25}
\end{figure}

Fig.\ \ref{fig:correctedDetProbs25} shows the corrections of Eq.\ \ref{eq:afterpulseCorr} applied to a typical $p_\mathrm{click}$ data set.  Note that bin 2 consistently experiences the greatest correction as a consequence of afterpulsing following the significant number of detection events in the first time bin.  With $\eta_d = 10$\%, $p_a$ was measured at a modest $\sim3$\%.  Given our inability to presently account for afterpulsing in the signature based reconstruction analysis, $\eta_d = 10$\% was chosen for all subsequent experimental data runs.

\section{Analysis of results}
\label{sec:analyseLoop}

In order to evaluate the proposed signature-based reconstruction method, a number of coherent state input tests were performed.  $1.5\times 10^5$ fixed amplitude pulses were acquired by the loop detector for each of a range of calibrated input intensities.  The data was then processed to form two probability distributions - $\hat{p}_\mathrm{click}(i)$ for time bins (and thus effective $\eta$) $i=1\ldots 10$ and $\hat{p}_\mathrm{sig}(\vec{d})_{N=9}$ for all $2^N = 512$ signatures.

The photostatistics of the input states were assumed to be Poissonian.  For each input amplitude, the MSE given by Eq.\ \ref{eq:mseTomog} was calculated from the experimental data set and a range of predicted data sets generated by Eq.\ \ref{eq:pClickBinom} and Eq.\ \ref{eq:signaturesTomog} for the binomial and signature schemes respectively.  To create the predicted data sets, a spectrum of estimated density matricies were generated with Poisson-distributed photon numbers where $\bar{n}$ was swept over the interval $[0,\sim15]$ photons/pulse.  Each of the evaluated input states were renomalised for a unity trace to account for the truncated Hilbert space, with the diagonal coefficients thus given by

\begin{equation}
\label{eq:coherentVals}
\rho_{nn} = \frac{\bar{n}^n}{n! \sum_{i=1}^K \frac{\bar{n}^i}{i!}}
\end{equation}

\begin{figure}[!htb]
\begin{center}
    \includegraphics[width=0.5\textwidth]{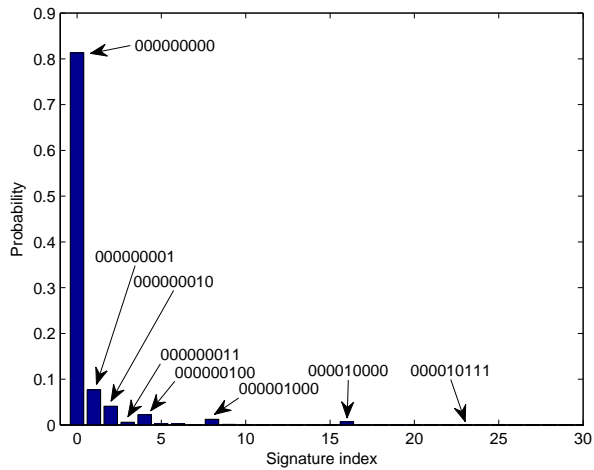}
    \caption{Predicted 9-bit signature probabilities for $\eta_d = 10$\%, $\bar{n}=8$ photons/pulse.}
    \label{fig:sigProbs}
\end{center}
\end{figure}

By means of illustration, Fig.\ \ref{fig:sigProbs} shows the predicted magnitudes of the first 30 $\hat{p}_\mathrm{sig}$ probabilities for an input coherent state corresponding to $\bar{n} = 8$ photons/pulse.  These probability values were then compared to the measured experimental probabilities via Eq.\ \ref{eq:mseTomog} to determine the associated MSE.  The indicated binary signature vectors correspond to the base-10 signature index, stored in bin-reversed order, i.e. time bin 1 is the right most bit \footnote{This is a consequence of the endianness of the FPGA FSM and has no practical significance}.  For example, the signature vector corresponding to a detection in the first and third time-bins is $\vec{d} = \{0,0,0,0,0,0,1,0,1\}$. This may be represented as a base-2 (binary) number, i.e. $000000101_2$, which has an equivalent base-10 (decimal) signature index $5_{10}$.  Signature index $6_{10}$ thus corresponds to $m=2$ detection events -- one in each of the second and third time bins and so on.  As expected from the $P(m|n)$ data of Section\ \ref{sec:loopChar}, the probability of multiphoton detection events decreases rapidly with increasing $m$, hence the tiny value shown for index $23_{10} = 000010111_2$, with $m=4$.

\begin{figure}[!htb]
\begin{center}
    \includegraphics[width=0.5\textwidth]{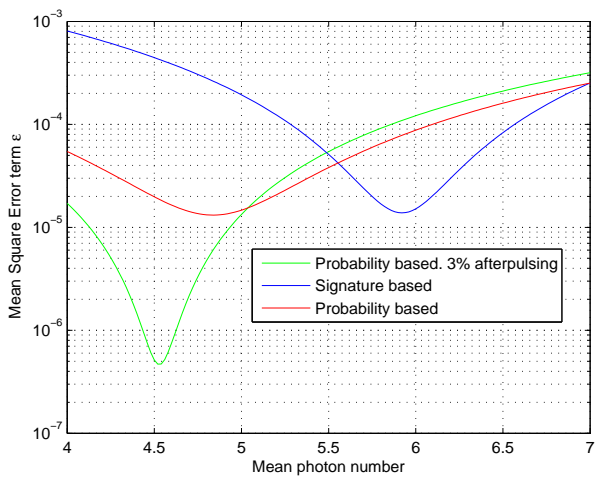}
    \caption{Mean-square-error for coherent-state fits to raw, corrected and signature probability data for $\bar{n}=6.5\pm0.8$ photons/pulse.}
    \label{fig:MSEfits}
\end{center}
\end{figure}

Fig.\ \ref{fig:MSEfits} illustrates the calculated values for $\epsilon$ against mean photon-number $\bar{n}$ for the experimental binomial $\hat{p}_\mathrm{click}$ and $\hat{p}_\mathrm{sig}$ signature probabilities for an input state of $\bar{n}=6.5\pm0.8$ photons/pulse.  A third trace illustrating the results of considering the $\hat{p}_\mathrm{click}$ data corrected for afterpulsing as per Eq. \ \ref{eq:afterpulseCorr} is also shown.  Note that the latter trace exhibits a very much smaller minimum value of $\epsilon$. This suggests a much closer agreement between the modelled and detected data, and the hence the validity of the applied correction.  The minimum value of $\epsilon$ for a given reconstruction scheme indicates the value of $\bar{n}$ that results in optimum agreement between the predicted and experimentally observed detection probabilities.  The offsets between the minima of the three curves result from the inherent sensitivity of the binomial approach to the absolute values of $\eta$ used in Eq.\ \ref{eq:pClickBinom}.  While approaches to self-calibration have been proposed to mitigate these effects \cite{AchillesPRL,BondaniArXiv}, the signature reconstruction method appears relatively insensitive to variations of individual values of $\eta$.  For example, 5\% variations of the switch loss term $t_s$ shifted the signature minimum by a corresponding amount, yet induced $>10$\% shifts in the binomial $\hat{p}_\mathrm{click}$ minimum.  Note also that the signature $\epsilon$ value is a more sharply peaked function, potentially improving the accuracy of photon number estimation.

\begin{figure}[!htb]
\begin{minipage}{0.45\linewidth}
\raisebox{-1ex}{\makebox[\textwidth][l]{\textbf{a)}}}
\centering
\includegraphics[width=\textwidth]{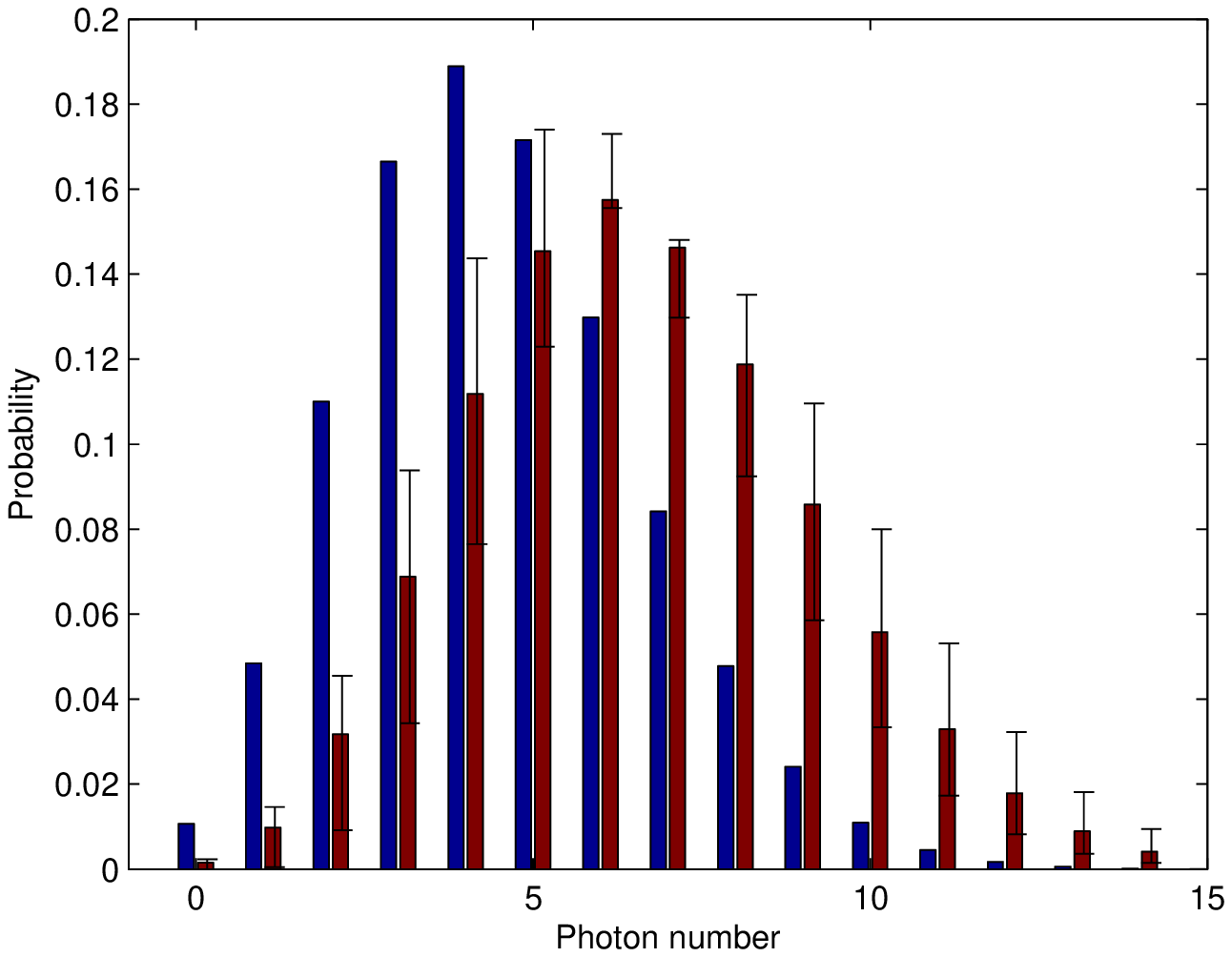}
\end{minipage}
\begin{minipage}{0.45\linewidth}
\raisebox{-1ex}{\makebox[\textwidth][l]{\textbf{b)}}}
\centering
\includegraphics[width=\textwidth]{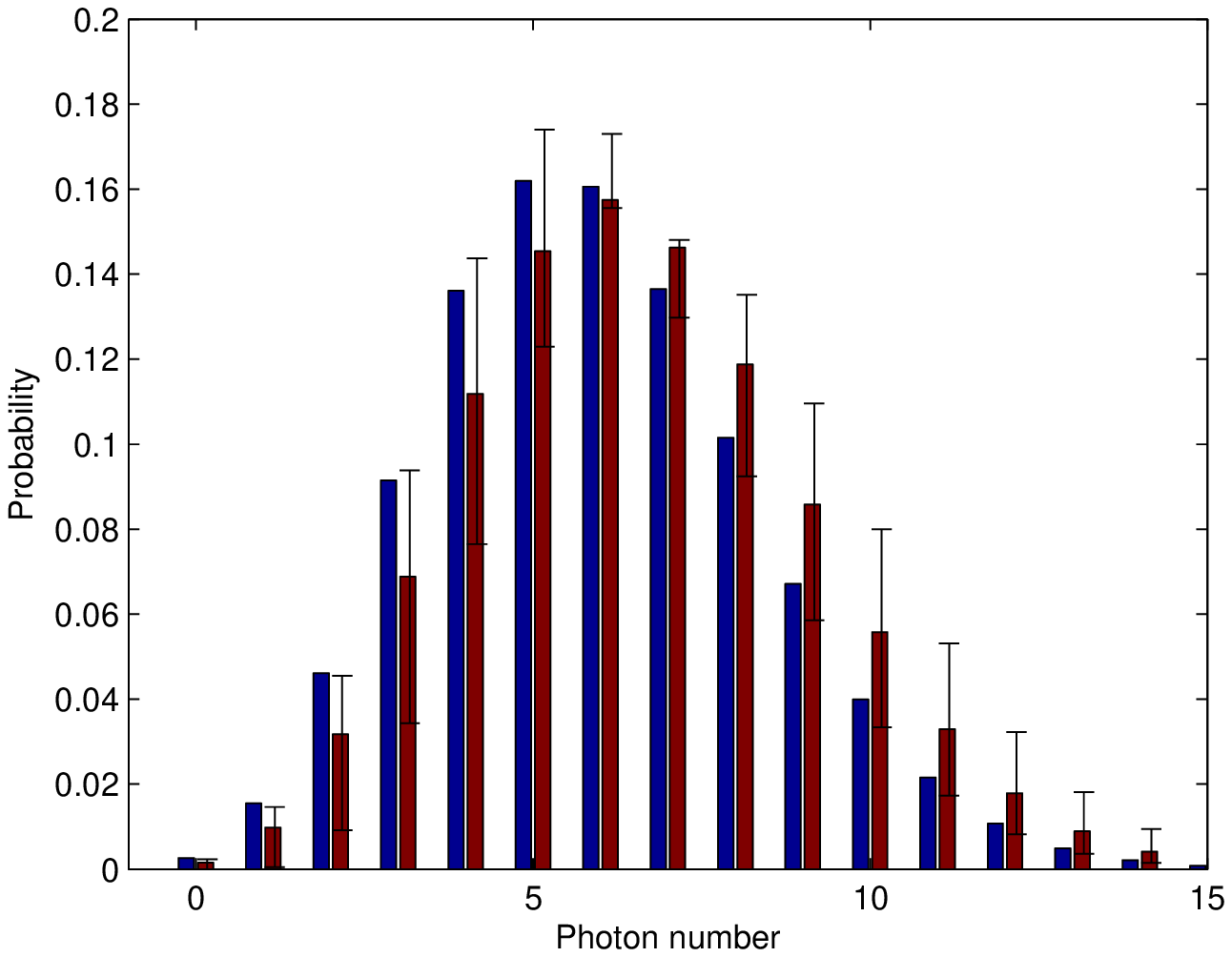}
\end{minipage}
\caption{Photon-number distributions corresponding to the observed $\epsilon$ (MSE) minima of Fig.\ \ref{fig:MSEfits}.  The best fit to the a) corrected binomial data occurs with $\bar{n}=4.54$ photons/pulse, and the b) signature data at $\bar{n}=5.95$ photons/pulse.  The photostatistics (and derived error bounds) of the $\bar{n}=6.5\pm0.8$ photons/pulse. input state is shown for comparison.}
\label{fig:pndistErrBar}
\end{figure}

The significance of the departures of the estimated values of $\bar{n}$ from the actual input state is examined in Fig.\ \ref{fig:pndistErrBar}.  Fig.\ \ref{fig:pndistErrBar} shows the photon number distributions corresponding to the values of $\bar{n}$ identified by the (afterpulsing-corrected) binomial and signature approaches of Fig.\ \ref{fig:MSEfits}.  For each case, the input distribution is shown for comparison.  The error bars were determined from the calibrated experimental uncertainty of the input amplitude.  The severe consequences of the calibration-induced errors of the binomial scheme are starkly visible, with the estimated distribution significantly different from the input state.  Conversely, the estimated signature statistics closely agree with the actual input statistics.

\begin{figure}[!htb]
\begin{center}
    \includegraphics[width=0.5\textwidth]{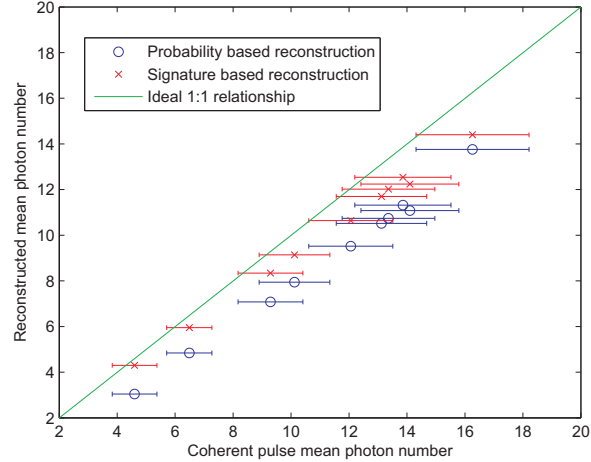}
    \caption{Estimated mean photon numbers for binomial probability and signature based reconstruction techniques.}
    \label{fig:reconsPN}
\end{center}
\end{figure}

Fig.\ \ref{fig:reconsPN} indicates the inferred photon numbers corresponding to the minima of the MSE $\epsilon$ error term for the two approaches over a range of input amplitudes.  The ideal 1:1 relationship is also shown.  The signature approach consistently returns values within the error bounds of loop detector losses and laser intensity calibration.   The linearly increasing error corresponds to a systematic calibration error of $\pm15$\% or 0.6dB, readily accounted for in the multiple insertion loss uncertainties arising from fibre-fibre interconnections.

\begin{figure}[!htb]
\begin{minipage}{0.45\linewidth}
\raisebox{-1ex}{\makebox[\textwidth][l]{\textbf{a)}}}
\centering
\includegraphics[width=\textwidth]{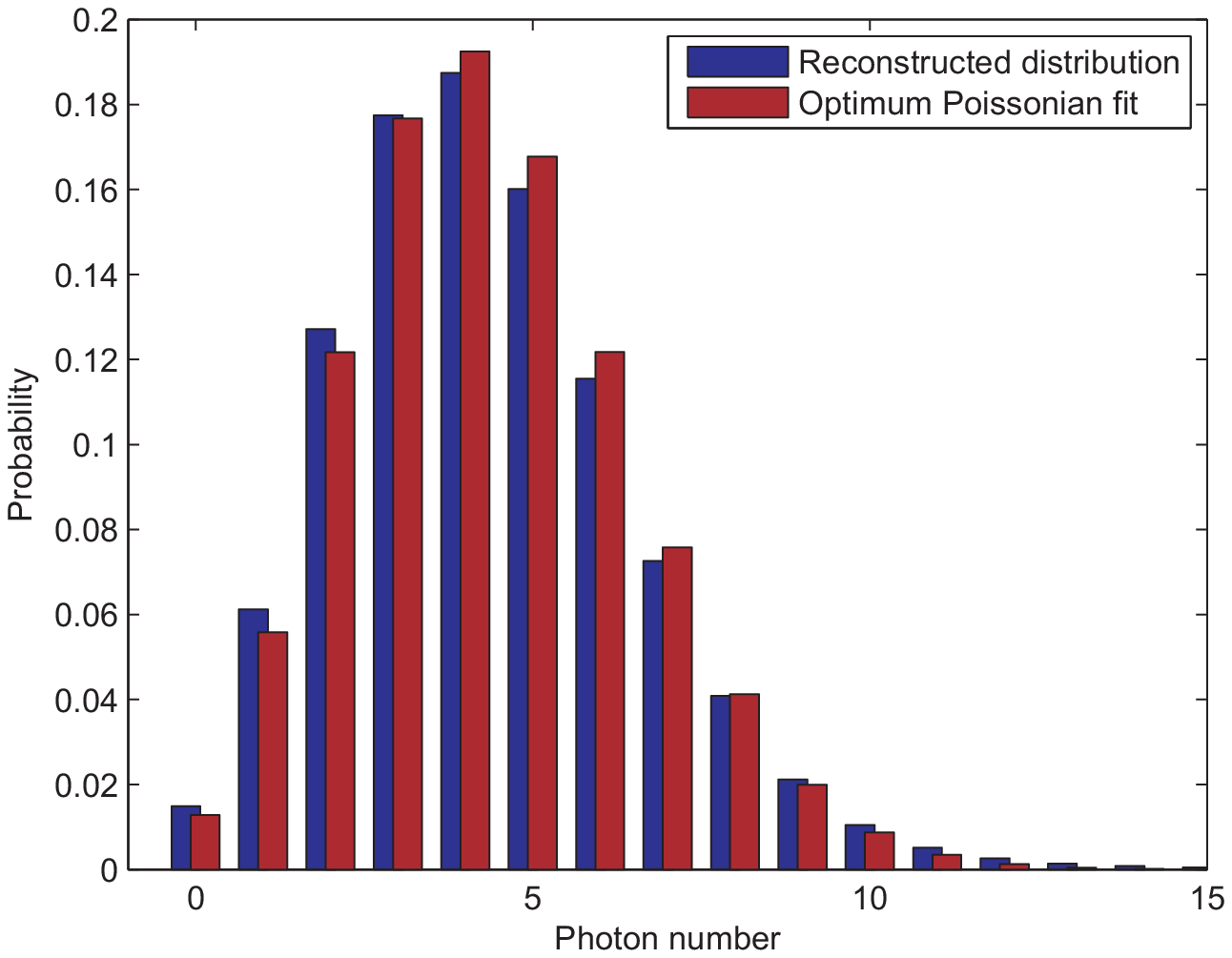}
\end{minipage}
\begin{minipage}{0.45\linewidth}
\raisebox{-1ex}{\makebox[\textwidth][l]{\textbf{b)}}}
\centering
\includegraphics[width=\textwidth]{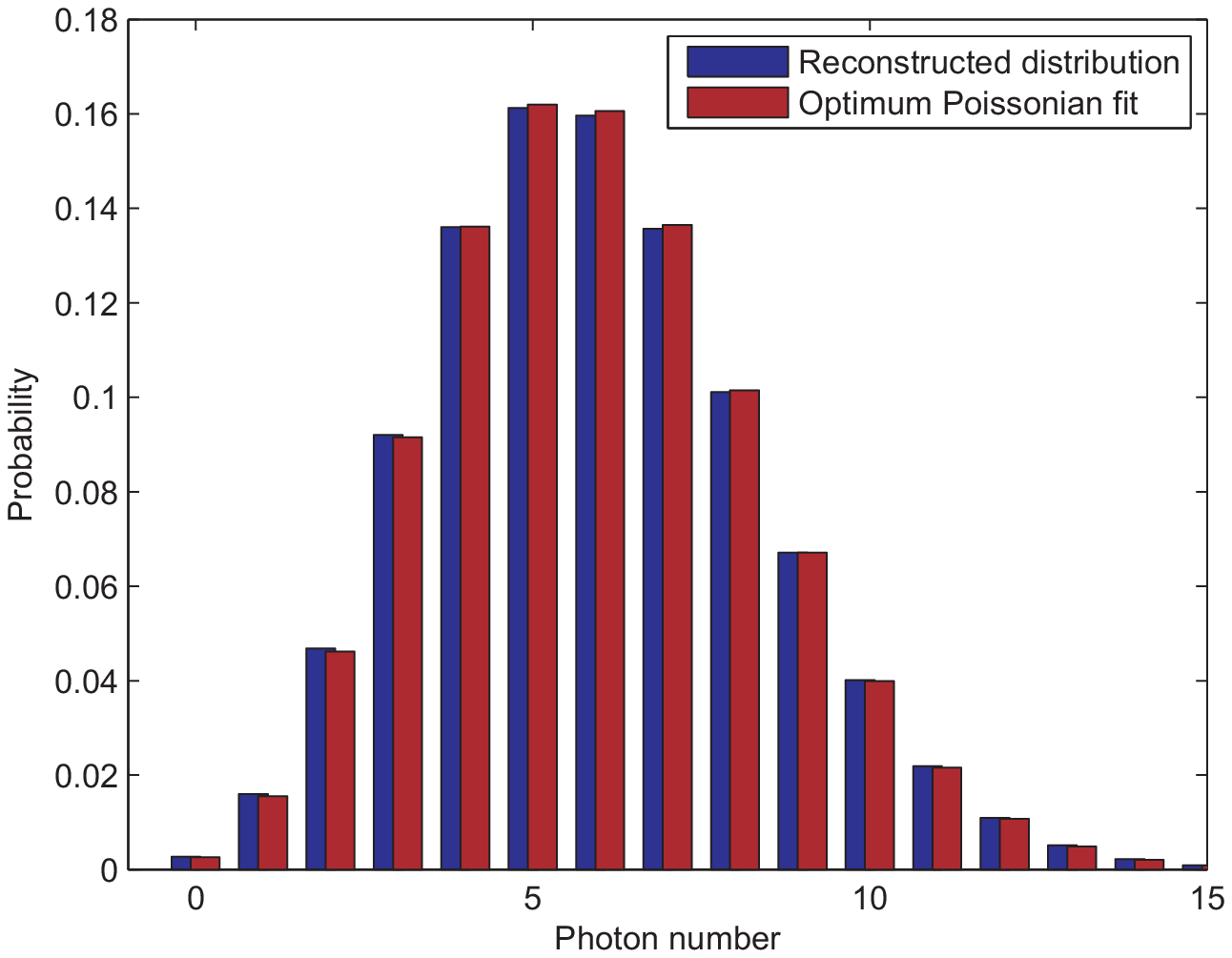}
\end{minipage}
\caption{Reconstructed photon-number distributions for an input coherent state with a) $\bar{n}_i = 4.6\pm0.8$ and b) $6.5\pm0.8$ photons/pulse.  Poissonian distributions representing the same mean photon number as the reconstructed states ($\bar{n}_r = 4.3$ and $6.0$ photons/pulse. respectively) are shown for comparison.}
\label{fig:optReconstruction}
\end{figure}

Finally, we investigate the performance of the signature reconstruction approach when no a priori assumptions are made regarding the form of the input photostatistics.  In this case, we seek to find the density matrix $\rho$ that minimises the MSE error $\epsilon$ by searching $(K+1)$-dimension parameter space for an optimum set of diagonal coefficients.  While we make use of the same coherent state experimental data used in the analysis above, the diagonal photon-number terms of the reconstructed density matrix are considered independent from one another and not bound to assume Poissonian form.  To this end, we employed the numerical multidimensional simplex minimisation algorithm described in Section\ \ref{sec:numMethods} to determine an optimal estimate of $\rho$ that minimises $\epsilon$. The expression for MSE shown in Eq.\ \ref{eq:mseTomog} was modified to incorporate unity trace constraints

\begin{equation}
\label{eq:mseTomogFinal}
\epsilon = \big(\mathrm{Tr}[\rho] - 1\big)^2 + \sum_{\vec{d}} (\hat{p}_\mathrm{sig}(\vec{d}, \hat{\rho})_{N=9} - p_\mathrm{sig}(\vec{d}, \rho)_{N=9})^2
\end{equation}

\noindent in a manner that makes no other assumption about the physicality of the estimated state.  The convergence properties of this expression in this application have not been comprehensively investigated.  Similarly, Eq.\ \ref{eq:mseTomogFinal} makes no attempt at optimisation of the likelihood or entropy of the reconstructed state \cite{HradilRehacek}.  This is identified as an area for further investigation.  Nevertheless, the results of reconstruction of the input state photostatistics with a) $\bar{n}_i = 4.6\pm0.8$ and b) $6.5\pm0.8$ photons/pulse. are quite impressive, as shown in Fig.\ \ref{fig:optReconstruction}.

To permit validation of the reconstructed distributions, the mean photon numbers of the reconstructed density matrices were determined via $\bar{n} = \sum_{n=0}^K n\rho_{nn}$.  Poissonian distributions with the corresponding mean photon numbers are shown beside the reconstructed distributions of Fig.\ \ref{fig:optReconstruction} for comparative purposes.  The agreement between the coefficients is remarkably close, confirming the practical suitability of the proposed signature approach for the reconstruction of photostatistics.

\section{Conclusion}
We have the described the use of photon count signatures from an N-port photon counting device for the purposes of reconstructing the photon number probability distribution of an input state, and demonstrated this technique experimentally on a 9-port system.   We have shown that exploiting signature probabilities offers superior performance to approaches reconstructing photon number probabilities from the measured binomial statistics.  Our method offers advantages in reduced statistical uncertainty, relative insensitivity to calibration errors and reduced reliance on {\em a-priori} assumptions about the input state.

\section*{Acknowledgments} This work was supported by the Australian Research Council Centres of Excellence scheme.


\end{document}